# THE FOURTH 4TH INDUSTRIAL REVOLUTION'S EFFECT ON THE ENTERPRISE CYBER STRATEGY

Christopher L. Gorham

Capitol Technology University

## ABSTRACT

*The Fourth (4th) Industrial Revolution represents the profound advancement of technology that will likely transform the boundaries between the digital and physical worlds in modern society. The impact of advance technology will disrupt almost every aspect of business and government communities alike. In the past few years, the advancement of information technologies has opened the door to artificial intelligence (AI), block chain technologies, robotics, virtual reality and the possibility of quantum computing being released in the commercial sector. The use of these innovative technologies will likely impact society by leveraging modern technological platforms such as cloud computing and AI. This also includes the release of 5G network technologies by Internet Service Providers (ISP) beginning in 2019. Networks that rely upon 5G technologies in combination with cloud computing platforms will open the door allow greater innovations and change the nature of how work is performed in the 4th Industrial Revolution.*

## INTRODUCTION

The 4th Industrial Revolution is upon the world today due to the technological advancements that has impacted businesses and governments across the globe. However, there are divergent opinions on how the potential effects will impact the workforce and the skills gap that has caused difficulties for many businesses in finding skilled qualified hiring talent. Andrew Johnson, writer and author, presents in his article titled[1] "Digital Skills are the Key to a higher paying job" that some employers are facing skill shortages and choose to focus on upskilling their own employees to address that gap. This strategy assumes that every employee that accepts any training to address a skills gap will be able transition easier into jobs that requires more technical knowledge and skills. But there are some drawbacks to this approach because not all employees have the resources to adapt to newer technologies that would allow them to perform their jobs. In addition, employers' effort to upskill their workforce would most likely require a significant investment that some smaller businesses probably cannot afford as opposed to larger organizations. This represents a dilemma for a portion of the workforce who works for an employer that lacks the resources to keep up with modern technologies advancement. The lack of resources that would limit their ability to upskill their employees would have a negative impact on their effectiveness towards innovation. The other effect would be larger employers that have the resources to upskills their employees would consume the marketplace.

Johnson argues that businesses need to have a collective focus on building technology skills for their existing workforce[1]. Businesses that heavily rely on technology to make production more efficient will need to have employees develop modern technology skills for the organization to have success and benefit from this approach. Employees who also work these large businesses will see benefits as well in attracting larger salaries that will allow them to provide more income towards supporting their families and lifestyle. Johnson makes a point in emphasizing the best way to grow the economy is raising the productivity level of organizations[1]. By employers investing resources into training their employees to adopt modern technology skills, they can be better prepared to innovative in this new technological age.

**U.S. COMPANIES HIRING FOREIGN TECH-WORKERS**

Over the past decade, U.S. technology companies such as Amazon, Microsoft and Apple have turned to hiring foreign workers through a visa program called H-1B. Through the H-1B program, companies are able to hire foreign workers in fields such as Science, Technology, Education and Math which is also referred to as STEM. Author Zoe Bernard writes in an article titled "Amazon is Hiring More Skills Immigrant…H-1B workers", that the number of applicants for H-1B work visa has exceeded the program's cap at 85,000 for the last 15 years [2].The demand for more Science, Technology, Education and Math (STEM) workers from tech companies has increased because of the advancement of technology from the products these companies provide to consumers. Amazon, Microsoft and Google make up the majority of companies that rely upon the H-1B visa worker program to hire skilled technology talent. The trend has increased over the past couple of years with Amazon requesting the most applicants [2].What this trend suggest is the reliance on technology has created a greater demand from employers to hire workers who are highly skilled in information technology. However, the fact that companies such as Microsoft, Apple and Google have to hire foreign workers with these skills indicates they can't find skilled workers in the United States. Despite the investment from businesses in training their own workers, the demand for more skilled technology employees is so great that they must seek talent outside the country.

The plan to hire more foreign tech workers from abroad will become even more difficult to accomplish because of the U.S. government's plan to limit H-1B visas. Rani Mollaexplains in an article written from the Vox news titled "Visa approvals for Tech Workers Are on the Decline",that the U.S. government is working to limit the amount of H-1B work visa to U.S. companies[3]. The assumption the article makes is that these businesses affected by this policy change will move their jobs overseas [3]. The author states this will have a negative effect on the U.S. job market for workers seeking high paying tech jobs to support their families. As the U.S. government stands firm on this policy, Stuart Anderson of the National Foundation for American Policy, says that restricting foreign nationals from obtaining H-1B visa displays a lack of understanding of how important IT services are for U.S. companies [3].

The chart by Statista (see below) shows that Amazon, Google and Microsoft represent a large portion of technology companies that rely upon H-1B visa workers [2]. The data represents an increasing trend by these companies that displays approved applications for H-1B working visas rose from 2016 to 2017 [2]. Amazon had the most applications during this time as the reliance on foreign workers to fill-in the need for technology talent has increased over the past few years.

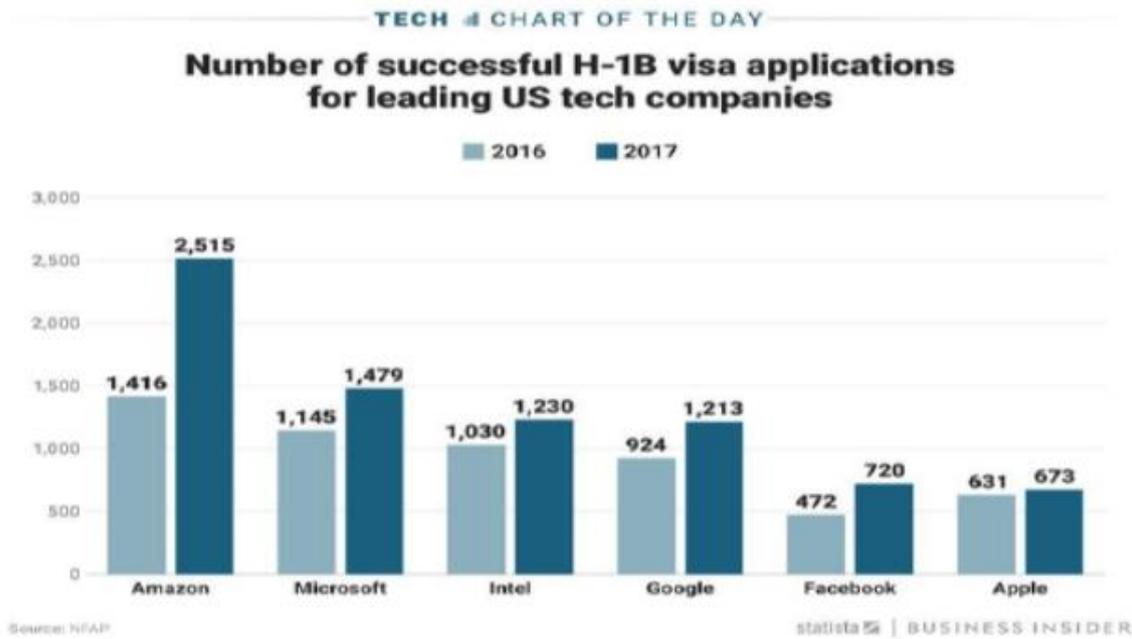

Figure 1. Number of successful H-1B visa applications

The chart provided by Recode (see below) displays the effect of the U.S. government's decision to limit H-1B visas is having a negative impact on their ability to hire top technology workers [3]. The approval rates for H-1B visas have gone down and almost 60% of applications now require extensive paperwork to justify approval for them[3]. In the past, a limited supply of H-1B workers may have resulted in less jobs and lower wages for workers in the US [3]. But the negative side effects are notable because businesses rely upon foreign workers on these programs to help them innovate, grow and employ more workers in high demand technology jobs [3]. As the job market becomes tighter for major technology companies, the H-1B visas worker program has become a source of relief to fill the gaps in the workforce [3]. However, as the demand for highly skilled technology workers becomes important, the competition for H-1B visa approvals has become harder to achieve due to the policy changes affecting the program [3].

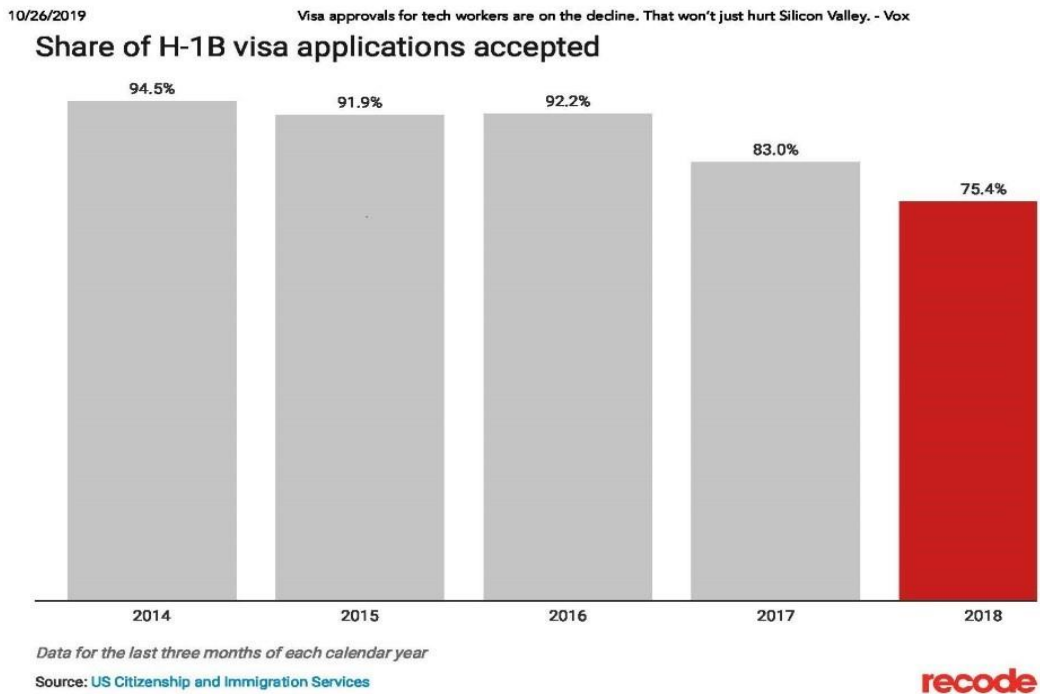

Figure 2. Share of H-1B visa applications accepted

## IMPACT ON THE U.S. DEPARTMENT OF DEFENSE

The U.S. policy towards cutting H-1B worker visas for foreigners may appear to be beneficial to U.S. workers but the skills gap problem still remains a potent issue. If not enough U.S. workers have the skills to compete for highly qualified technology jobs, then it raises a question of how companies will be able compete for skilled talented workers. This will have a dramatic effect on the Department of Defense's ability to upgrade their network to meet today's technological standard. DOD will be relying upon technology companies such as Apple, Microsoft, Amazon and Google to help them update their network infrastructure. Nevertheless, if the current administration's policy towards limiting these companies' ability to hire talent through the H-1B visa program financially impacts them then how can they succeed to meet this demand? The answer is probably not as simple for a company that has a history of valuable services to help an organization meet their mission goals. But it's when the negative effect of a policy change makes it complicated while limiting an organization's ability to perform the services necessary that makes them successful that will cause the greater impact. In all likelihood, it's possible that DOD could create other ways of hiring more homegrown talent to make up for the loss. It is questionable that they could succeed if they are unable to find the tech talent they need to meet their organizational goals [2].

DOD is taking steps towards addressing the need of providing a skilled capable technology workforce [2].In their budget proposal to the U.S. Congress, DOD requested funding for a new senior level IT position that focuses on recruiting and promoting a digitized workforce. As argued by Jackson Barnett writes in an article presented by FedScoop titled "Congress wants to give DOD a top IT workforce recruiting official" that this position will be mandated to "track the digital readiness of the military and civilian workforce across DOD". Since this proposed position will be at the level of the Assistant Defense Secretary, it will allow this official to provide the DOD leadership an inside assessment of the current workforce capabilities while DOD is focusing on transitioning into the modern technological era [2].Whether this approach changes the nature of DOD's workforce to enhance their technological capabilities remains questionable at best. The effectiveness of this plan will be dependent upon DOD's ability to recruit highly skilled workers that offers flexibility into positions that influences positive changes.

The emergence of AI and other advanced technologies will force organizations to adapt the skillset of their workforce so they can adjust to the faster pace of technological change[4]. DOD will need to develop a program that allows their workforce to compete for jobs that require a greater skillset [2].As the need for innovative solutions becomes greater in the era of the 4th Industrial Revolution, organizations will need to response to the technological challenges the world faces today. It's therefore essential that DOD stays committed towards building a modern technology workforce to help them prepare for future cyber challenges [2].

## SHORTAGE OF STEM JOBS FOR STEM GRADUATES

If DOD is going to take steps to retrain or hire highly skilled workers for their workforce, they will need to invest into STEM (Science, Technology, Engineering and Math). If not enough U.S. workers have the skills to compete for highly qualified technology jobs then it raises a question of how companies will be able compete for skilled talented workers. Another issue that may complicate matters is the possible shortage of STEM jobs for STEM graduates. This issue presents a dilemma to the concept that thereisn't enough workers for STEM jobs according author Roberta Rincon, Phd. Rincon writes in an article titled "Is There a Shortage of STEM Jobs to STEM Graduates"that most of the debate over the shortage of STEM jobs or STEM graduates centers on how the U.S. government defines "STEM" itself (Rincon, 2019). She presents data to support her argument that showed a graph (see below) has more graduates for engineers than the actual demand [5]. If this is the case, Rincon concludes that more efforts need to be made to encourage underrepresented groups to complete engineering degrees to enter the profession to compete for those jobs [5].

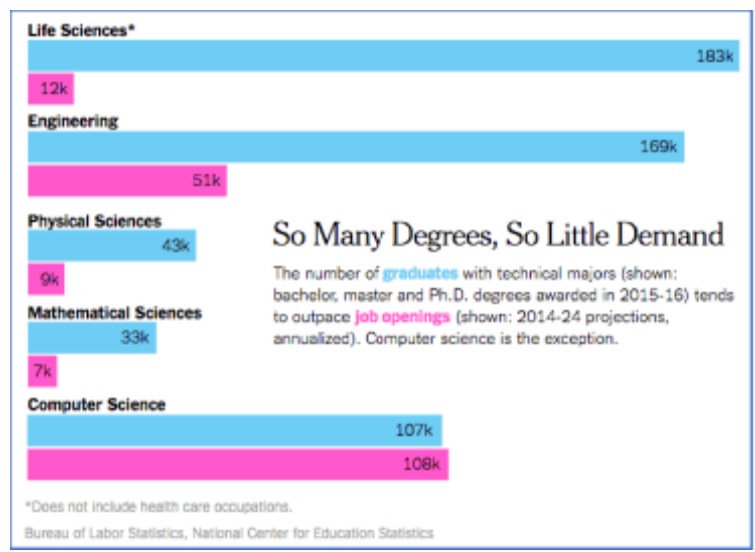

Figure 3.  Number of college graduates with technical majors

## H-1B IMPACT ON DOD'S IT CONTRACTOR COMPANIES

The success or failure of STEM in the United States will have a tremendous impact on the workforce in the U.S. due to the need for advanced technological skillsets. But if the level of STEM graduates don't meet the demands of the job market then the normally reliable H-1B program represents a backdrop of finding talent for STEM jobs. In an article titled "IT Consulting Industry Hardest Hit by Jump in H-1B Denials", Laura Francis explains while temporary visas in the U.S. have increased, IT consulting companies have been negatively impacted recently[6].Francis points out that the IT consulting industry "saw lower approval rates" for their request for H-1B visas even though some of the appeals to those decisions were still approved [6]. This is an important factor for organizations like DOD because consultants are able to advise clients on how to best move forward with enterprise level IT projects. With DOD looking to upgrade their cyber infrastructure by hiring an IT workforce recruitment official to lead their efforts, having consults can have a meaningful purpose[7]. As a results of these studies, the impact could potentially cause delay for DOD in implementing efforts to update their network if they can't hire talent via the H-1B worker program. The combination of less STEM graduates and reduction of H-1B could be an obstacle too much to overcome if no solution is provided to compete in this new era of the 4th Industrial Technological Revolution [4].

## SOLUTIONS

One of the many solutions being discussed is for companies to take an approach to solve the issue of skills gap from the inside.In an article titled "The US has Nearly 1 Million Open Jobs", Jennifer Liu writes that companies can bridge the tech gap by retraining their current employees[8].Liu lists examples of people who successfully transitioned from other jobs into tech related jobs [8]. This provides a small template of success for organizations to follow if they invest in internal training for their employees [8]. While this solution may benefit companies they have resources to conduct in-house training, it remains of question of commitment if they're able to transition employees to positions with the right skillsets. This strategy will certainly benefit organizations such as DOD if they are unable to reverse the trend of reducing the government's reliance on H-1B visas[9]. It provides another to consider option until a more effective approach has been offered.

In an article titled "The Digital Skills Gap is widening fast, Here's how to bridge it",Miguel Milano statesthat a viable short-term solution to bridging the skills gap is for businesses to acquire premium talent [10]. By working with staffing agencies and recruiters, companies will be able to find qualified skilled talent with experience that fits the needs of the organization. However, Milano dismisses this approach as a longer term solution because newer skills will be in demand [10]. The author instead recommends a multi-tier approach with several options to solve the skills gap [10]. The recommendation is for businesses to prioritize skills development "within their existing workforce" along talent that maybe untapped but has potential [10]. This task can be accomplished by allowing hiring managers to setup workforce development programs to help the workforce better prepare for technological innovations. These programs will need to be integrated into organization's operational costs as a permanent expense so they aren't held back by budgetary constraints [10]. Milano also suggests that businesses use technology to enable lifelong learning programs that have courses for employees so they can reskill or upskill their abilities to adapt to the changing nature of the workforce [10]. According to Milano, this plan will allow employees to take control of their own training plan by developing skillset they need to keep up with technological changes [10].

Lastly, Milano suggest organizations will be better served looking for talent outside of the traditional talent pools [10]. This means taking a more progressive approach towards hiring skilled talent from people with diverse backgrounds and those without a college degree [10]. Businesses will gain more skilled workers by opening up the hiring process to more people from different backgrounds and developing that talent using the programs Milano had suggested. This strategy will take a wholesome commitment from business leaders to diverge from the traditional approaches in the past. But it will also open the door for businesses and the workforce to stay up to the date in the modern technology world that will change the very nature of work itself.

## CONCLUSION

The immediate effect of the 4th Industrial Revolution on society will potentially change the way it operates in every area of business, commerce, communications and government. The advancement of technology will probably blur the lines between the digital and physical worlds as more businesses adopt modern technological tools such as AI and robotics. How much impact these changes will have will largely depend on how disrupted these technologies can be in the business and government sectors. As technology becomes more innovative, businesses will need to decide how to make decisions that will affect their workforce and the required skills they need for the new jobs that are being created. Government agencies such as DOD will need a strategic plan on how to create, manage and defend their cyber networks in an era that provides both law enforcement, military and cyber criminals tools to exploit technology to their advantage. To accomplish those goals, building a network that leverages modern technological platforms such as cloud computing will provide the flexibility they need for these innovative tools to have its greatest effect as they implement their strategy to move forward.